\documentclass[12pt]{article}
\usepackage{amsmath,amssymb,epsfig}

\usepackage{color}
\input{colordvi.tex}

\makeatletter

\renewcommand\section{\@startsection {section}{1}{\z@}%
                                   {-3.5ex \@plus -1ex \@minus -.2ex}%
                                   {2.3ex \@plus.2ex}%
                                   {\normalfont\large\bfseries}}

\renewcommand\subsection{\@startsection{subsection}{2}{\z@}%
                                     {-3.25ex\@plus -1ex \@minus -.2ex}%
                                     {1.5ex \@plus .2ex}%
                                     {\normalfont\normalsize\bfseries}}

\newcount\hour \newcount\minute
\hour=\time \divide \hour by 60
\minute=\time
\def\now{%
\ifnum \hour<13
  \ifnum \hour=0 \advance \hour by 12 \number\hour:\else \number\hour:\fi%
     \ifnum \minute<10 0\fi%
     \number\minute%
\ A.M.%
\else \advance \hour by -12 \number\hour:%
  \ifnum \minute<10 0\fi%
  \number\minute%
  \ P.M.%
\fi%
}

\makeatother

\begin{document}

\baselineskip=18pt  
\numberwithin{equation}{section}  
\allowdisplaybreaks  



%
%


\thispagestyle{empty}

\vspace*{-2cm}
\begin{flushright}
YITP-10-29\\
\end{flushright}

\begin{center}

\vspace{3cm}

{\bf\Large Inflation and Gauge Mediation
\vspace{0.3cm}  \\  in Supersymmetric Gauge Theory}
\\

\vspace*{2.0cm}
{\bf
Yuichiro Nakai$^{\dagger,}$\footnote{E-mail:
ynakai@yukawa.kyoto-u.ac.jp} and Manabu Sakai$^{\dagger,}$}\footnote{E-mail:
msakai@yukawa.kyoto-u.ac.jp} \\
\vspace*{0.5cm}

$^{\dagger}${\it {Yukawa Institute for Theoretical Physics, Kyoto University,
   Kyoto 606-8502, Japan}}

\vspace*{0.5cm}

\end{center}

\vspace{1.5cm} \centerline{\bf Abstract} \vspace*{0.5cm}

We propose a simple high-scale inflationary scenario based on a phenomenologically viable model with direct gauge mediation of low-scale supersymmetry breaking. Hybrid inflation is occurred in a hidden supersymmetry breaking sector. Two hierarchical mass scales to reconcile both high-scale inflation and gauge mediation are necessary for the stability of the metastable supersymmetry breaking vacuum. Our scenario is also natural in light of the Landau pole problem of direct gauge mediation.

\newpage
\setcounter{page}{1} 





\section{Introduction}

Cosmological inflation in the early stage of our universe can solve a lot of problems existed in the standard Big Bang cosmology (for reviews, see \cite{Lyth:1998xn,Mazumdar:2010sa}). It can explain the flatness, horizon, and monopole problems. Furthermore, quantum fluctuations of the field in charge of inflation can set the initial condition of structure formation. Therefore, inflation is now considered as the standard scenario of the early universe.

Then, a natural question is how inflationary scenario fits in particle physics model building. It is known that a certain type of inflation, hybrid inflation\footnote{Hybrid inflation often suffers from unwanted topological defects. In this letter, we aim to propose a simple toy model to reconcile both inflation and gauge mediation.}, is naturally realized in supersymmetric theories \cite{Dvali:1994ms,Dimopoulos:1997fv,Craig:2008tv}. Supersymmetry (SUSY) is widely considered to play a fundamental role in the framework beyond the standard model of particle physics. Its minimal realization is called minimal supersymmetric standard model (MSSM).\footnote{There are attempts to realize inflation in the MSSM sector \cite{Allahverdi:2006iq,Allahverdi:2006we}, although they seem to require some fine-tunings \cite{Bueno Sanchez:2006xk,Kamada:2009hy}.} However, supersymmetry must be broken at an energy scale. It is usually assumed that SUSY breaking in the MSSM sector is transmitted from a hidden SUSY breaking sector by some interactions. Among many possibilities of the interactions, gauge mediation of SUSY breaking has several distinguished features (for reviews, see \cite{Giudice:1998bp,Intriligator:2007cp,Kitano:2010fa}). For example, it naturally suppresses the unwanted flavor-changing processes, due to the flavor blindness of the standard model gauge interactions. 

Many gauge mediation models are already known. Among them, direct gauge mediation is the model that the flavor symmetries in a hidden SUSY breaking sector are weakly gauged to become the standard model gauge symmetries. In this class of gauge mediation models, it is known that the standard model gaugino masses often vanish at the leading order of SUSY breaking \cite{Izawa:1997gs}. Then, there is a little hierarchy between the gaugino masses and the scalar masses without careful tuning of parameters. If this hierarchy exists, we cannot obtain the order 1 TeV gaugino masses and scalar masses at the same time, which causes the standard model hierarchy problem again. Recently, Komargodski and Shih presented a theorem for this widely known phenomenon \cite{Komargodski:2009jf}, which says that we can obtain nonzero leading order gaugino masses when the pseudomoduli space of a SUSY breaking vacuum is not locally stable everywhere. It was the model proposed by Kitano, Ooguri, and Ookouchi (KOO) \cite{Kitano:2006xg} that had explicitly realized this fact to generate sizable gaugino masses (see also \cite{Giveon:2009yu}). Their model is based on the recent development of metastable SUSY breaking in supersymmetric QCD. That is, they take a deformed model of the Intriligator-Seiberg-Shih (ISS) model \cite{Intriligator:2006dd} as a hidden sector of direct gauge mediation. Since the vacuum considered in the KOO model is not the global minimum of the potential, it is necessary to impose several conditions on the model parameters in order to guarantee the vacuum stability. In particular, it is required that this massive SQCD model has flavors with two hierarchical mass scales.\footnote{For a general discussion about the requirement of several mass scales, see \cite{Giveon:2009yu}.}

In this letter, we propose a hybrid high-scale inflationary scenario motivated by the KOO model as a first step to reconcile both inflation and gauge mediation with a single hidden sector.\footnote{In contrast to the original KOO model, we consider the IR-free magnetic description of the model from the start for our purpose. If we considered the electric description of the theory above the Planck scale in our scenario, a serious trans-Planckian mass scale would appear. Therefore, our IR-free gauge theory must be completed by another Planck scale physics.} Hybrid inflation based on the ISS model has already been discussed in \cite{Craig:2008tv}. However, this scenario leads to the unacceptably large SUSY breaking scale for phenomenological applications in order to realize high-scale inflation. In contrast, as explained above, our model naturally has two hierarchical mass scales, and hence we can determine the higher mass scale such that the correct amplitude of density fluctuations can be obtained. The lower mass scale corresponds to the SUSY breaking scale to give the MSSM soft masses by direct gauge mediation. Our scenario also has no Landau pole problem often encountered in direct gauge mediation.\footnote{For another approach to reconcile both inflation and gauge mediation in several hidden sectors, see \cite{Savoy:2007jb,Brax:2009yd}.}

The rest is organized as follows. In section 2, we will present the SUSY breaking model corresponding to the magnetic description of the KOO model and explain how two mass scales are required in this model. In section 3, we will present our inflationary scenario. In section 4, we will provide the inflationary predictions of our scenario, and show the parameter settings to give the order $1$ TeV gaugino and scalar masses. In section 5, we will give the conclusion with some discussions. Many more details of this scenario will be discussed in \cite{Kamada}.

\section{SUSY breaking}

The model is an $\mathcal{N}=1$ supersymmetric $SU(N)$ gauge theory with $N_f \, (>3N)$ flavors $q_i$, $\bar{q}_i$ $(i = 1, \cdots, N_f)$ of (anti-)fundamental representation and singlets $M_{ij}$. The tree-level K\"ahler potential is canonical, $K = |M_{ij}|^2 + |q_i|^2 + |\bar{q_i}|^2$. In addition, the renormalizable superpotential of the model is denoted as
\begin{equation}
W = m^2 M_{I I} + \mu^2 M_{a a} - h q_i M_{i j} \bar{q}_j - m_z M_{I a} M_{a I}, \label{magsp}
\end{equation}
where the indices $I ,a$ run for $I = 1, \cdots, N, \,\,\, a=N+1, \cdots, N_f$. The mass scales $m, \, \mu$ are assumed to be $m > \mu$ and much smaller than the cut-off scale of the theory which is set above the Planck scale in our inflationary scenario. $h$ is a small coupling constant, which we assume to be $\mathcal{O}(10^{-2})$.
$m_z$ is also a mass scale whose size is determined below.
The model has the global symmetry $SU(N) \times SU(N_f-N) \times U(1)_B$.

There exists a SUSY breaking vacuum, which we call the ISS vacuum, near the origin of the singlet field space. The F-term conditions for all components of $M_{ij}$ cannot be satisfied by the rank condition. The energy of that vacuum is given by $V_{ISS} = (N_f-N) \mu^4$, and the expectation values of the fields are denoted as
\begin{equation}
M_{i j} =
\begin{pmatrix}
0 & 0 \\
0 & \Phi_{ab}
\end{pmatrix}
, \quad
q_i = \bar{q}_i =
\begin{pmatrix}
\frac{m}{\sqrt{h}} \delta_{IJ}\\
0
\end{pmatrix},
\end{equation}
where $\Phi_{ab}$ are the pseudomoduli of the SUSY breaking vacuum. Their stabilized points at one-loop deviate from the origin by the amount of $\mathcal{O}(m_z)$. At the vacuum, the gauge group $SU(N)$ is completely broken. This vacuum is the destination end-point in our inflationary scenario.

In addition to the ISS vacuum, there are other vacua away from the origin which have lower energies than that of the ISS vacuum.\footnote{In the original KOO model, we also have the SUSY preserving vacuum generated by the non-perturbative dynamics. In our case, the cut-off scale is above the Planck scale, and hence the existence of this vacuum is unclear.} To see this, let us decompose singlet fields $M_{i j}$ and quarks $q_i , \bar{q}_i $ such as
\begin{equation}
M_{i j} =
\begin{pmatrix}
Y_{I J} & Z_{I a} \\
\bar{Z}_{a I} & \Phi_{ab}
\end{pmatrix}
, \quad
q_i =
\begin{pmatrix}
\chi_{I} \\
\rho_{a}
\end{pmatrix}
, \quad
\bar{q}_i =
\begin{pmatrix}
\bar{\chi}_{I} \\
\bar{\rho}_{a}
\end{pmatrix},
\end{equation}
and the superpotential \eqref{magsp} is rewritten by these component fields as
\begin{equation}\label{dualsuperpotential}
W = m^2 Y_{I I} + \mu^2 \Phi_{a a} - h \chi_I Y_{I J} \bar{\chi}_J - h \rho_a \Phi_{a b} \bar{\rho}_b - h \chi_I Z_{I a} \bar{\rho}_a - h \rho_a \bar{Z}_{a I} \bar{\chi}_I - m_z Z_{I a} \bar{Z}_{a I}.
\end{equation}
Then, we can find the other SUSY breaking vacua with energies $V_{low} = (N_f-N - n)\mu^4$, whose expectation values are given by
\begin{equation}
\begin{split}
\chi \bar{\chi} &= \frac{m^2}{h} \mathbf{1}_{N}, \quad \rho \bar{\rho} = \frac{\mu^2}{h} \mathrm{diag}(1 \cdots 1, 0 \cdots 0), \\
&\qquad Z \bar{Z} = \frac{m^2\mu^2}{m_z^2}\mathrm{diag}(1 \cdots 1, 0 \cdots 0),\\
Y &= \frac{\mu^2}{m_z} \mathbf{1}_{N}, \quad \Phi = \frac{m^2}{m_z} \mathrm{diag}(1 \cdots 1, 0 \cdots 0),
\end{split}
\end{equation}
where the number of non-vanishing diagonal components in the above expressions, denoted $n$, runs from $1$ to $N$. Since these vacua have lower energies, the ISS vacuum has a nonzero transition probability to these vacua. The decay rate can be estimated by evaluating the Euclidean action from the ISS vacuum to the lower vacua, which is obtained as
\begin{equation}
S \propto \Big( \frac{m}{\mu} \Big)^4.
\end{equation}
Therefore, the mass hierarchy $\mu \ll m$ is required in order to guarantee the stability of the ISS vacuum. This feature is crucial in our inflationary scenario as explained in the next section.

In order to transmit the SUSY breaking of the ISS vacuum to the MSSM sector by direct gauge mediation, we embed the standard model gauge group into the global symmetry $SU(N)$ of the model. The gaugino and scalar masses are given by
\begin{equation}
\begin{split}
m_\lambda &= \frac{g^2 \bar{N}}{16\pi^2} \frac{\mu^2}{m} \frac{m_z}{m} + O \left( \frac{m_z^2}{m^2} \right), \\
m_i^2 &= 2 \bar{N} C_2^i \left( \frac{g^2}{16\pi^2} \right)^2 \frac{h\mu^4}{m^2} + O \left( \frac{m_z^4}{m^4} \right),
\end{split}\label{soft}
\end{equation}
where $\bar{N} \equiv N_f-N$, and $g$ is the gauge coupling constant of the standard model gauge group. $C_2^i$ is a quadratic Casimir. As we can see in the above expressions, the leading order gaugino masses do not vanish. Since the ISS vacuum is the higher-energy state in the potential, this result is consistent with the general theorem presented in \cite{Komargodski:2009jf}. In order to obtain the same order gaugino and scalar masses, the parameter choice $m_z \sim m/\sqrt{\bar{N}}$ is required.

\section{The scenario}

We now consider the inflationary dynamics resulting from the above IR-free gauge theory with the superpotential \eqref{magsp}. As we will see below, hybrid inflation can be naturally occurred in the SUSY breaking sector. Singlet fields become inflaton fields, and quarks are waterfall fields of hybrid inflation.

To analyze the inflationary dynamics, we first parametrize the inflaton trajectory by 
\begin{equation}
M =
\begin{pmatrix}
\frac{\phi_1}{\sqrt{N}} \mathbf{1}_{N} & 0 \\
0 & \frac{\phi_2}{\sqrt{N_f-N}} \mathbf{1}_{N_f-N}
\end{pmatrix},
\quad q=\bar{q}=0,
\label{trajectory}
\end{equation}
so that it respects the $SU(N) \times SU(N_f-N) \times U(1)_B$ global symmetry of the model.
If we consider supergravity correction, as we will soon show explicitly, the $\phi_2$ field obtains a large mass of order $\mathcal{O}(m^4/M_{Pl}^2 )$, where $M_{Pl} = 1/\sqrt{8\pi G} = 2.4 \times 10^{18} \, \mathrm{GeV}$ is the reduced Planck mass scale. Thus we can fix $\phi_2$ at its vacuum expectation value on the ISS vacuum.
The general form of supergravity scalar potential is expressed as
\begin{equation}
V_{g} = e^{K/M_{Pl}^2} \left[ \left( \frac{\partial^2 K}{\partial \phi^\dagger_\alpha \partial \phi_\beta} \right)^{-1} \left( \frac{\partial W}{\partial \phi_\alpha} + \frac{W}{M_{Pl}^2} \frac{\partial K}{\partial \phi_\alpha} \right) \left( \frac{\partial W^\dagger}{\partial \phi^\dagger_\beta} + \frac{W^\dagger}{M_{Pl}^2} \frac{\partial K}{\partial \phi^\dagger_\beta} \right) - \frac{3}{M_{Pl}^2} |W|^2 \right].
\end{equation}
Along the inflationary trajectory \eqref{trajectory}, we obtain the following supergravity scalar potential, that is,
\begin{equation}
\begin{split}
V_g(\phi_1 ,\phi_2 ) &= N m^4 + (N_f-N) \mu^4 + \frac{(N_f-N) \mu^4}{M_{Pl}^2} \phi_1 \phi_1^\dagger + \frac{N m^4}{M_{Pl}^2} \phi_2 \phi_2^\dagger \\
&\qquad- \frac{\sqrt{N(N_f-N)}m^2 \mu^2}{M_{Pl}^2} (\phi_1 \phi_2^\dagger + \phi_1^\dagger \phi_2 ) + \cdots.
\end{split}
\end{equation}
Since the mass scale $\mu$ is small compared to $m$ as explained in the previous section, we can neglect the terms which contain $\mu$.
As we can see, only the field $\phi_2$ obtains a large mass $m_{\phi_2}^2 = \frac{N m^4}{M_{Pl}^2}$, which implies the slow-roll parameter for $\phi_2$ is an inappropriate value $\eta_2 \simeq 1$ for slow-roll inflation.

Then, we concentrate on the inflationary dynamics of the $\phi_1$ field. As far as the $\phi_1$ dynamics is concerned, we can ignore the supergravity corrections.\footnote{The non-canonical K\"ahler potential generated by loop effects might destroy our inflationary potential. However, when coupling constants in the present model are sufficiently small, it is considered that this kind of $\eta$ problem does not occur.} The tree-level scalar potential is given by
\begin{equation}
V_{tree} \simeq \left| m^2 \sqrt{N} - \frac{h}{\sqrt{N}} \chi \bar{\chi} \right|^2 + \frac{h^2}{N} |\phi_1|^2 ( |\chi|^2 + |\bar{\chi}|^2 ).
\end{equation}
At tree-level, the $\phi_1$ field still feels nothing for the potential, so inflation does not end. Let us calculate the quantum correction to the above potential by integrating out the massive fields $\chi, \,\bar{\chi}$. The masses of the scalar components along this trajectory are given by $m_{1s}^2 = \frac{h^2}{N}|\phi_1|^2 \pm hm^2$, while those of the fermions are $m_{1f}^2 = \frac{h^2}{N}|\phi_1|^2$. The one-loop effective potential for $\phi_1$ is then
\begin{equation}
\begin{split}
V_{eff}(\phi_1)
&= N m^4 + \frac{N^2}{32\pi^2 } \Biggl[ 2 h^2 m^4 \log{\left( \frac{h^2 \phi_1^2}{NM_{\ast}^2} \right)} \\
&\quad+ \left( \frac{h^2}{N} \phi_1^2 + hm^2 \right)^2 \log{\left( 1 + \frac{Nm^2}{h\phi_1^2} \right)} + \left( \frac{h^2}{N} \phi_1^2 - hm^2 \right)^2 \log{\left( 1 - \frac{Nm^2}{h\phi_1^2} \right)} \Biggr].
\end{split}
\end{equation}
$M_{\ast}$ denotes a cut-off scale. Due to this effective potential, the inflaton now rolls over. When $\phi_1 = \phi_1^c \equiv \sqrt{\frac{N}{h}}m$, the scalar fields $\chi, \,\bar{\chi}$ become tachyonic, and roll off to the ISS vacuum with nonzero vacuum expectation values, which ends the inflationary process.

In order to see that slow-roll inflation occurs properly, we need to know whether the slow-roll parameters satisfy the conditions $\epsilon, \eta \ll 1$. We now parametrize the inflaton trajectory by $x \equiv \frac{\phi_1}{\phi_1^c}$, which leads to the effective potential of the form,
\begin{equation}
\begin{split}
V_{eff}(x) &= N m^4 \Biggl[ 1 + \frac{N h^2}{32\pi^2 } \biggl[ 2 \log{\left( \frac{hm^2 x^2}{M_{\ast}^2} \right)} \\
&\quad+ (x^2 + 1 )^2 \log{\left( 1 + \frac{1}{x^2} \right)} + (x^2 - 1 )^2 \log{\left( 1 - \frac{1}{x^2} \right)} \biggr]\Biggr].
\end{split}
\end{equation}
Then, the slow-roll parameters are given by
\begin{equation}
\begin{split}
&\epsilon = \frac{M_{Pl}^2}{2} \left( \frac{1}{V} \frac{\partial V}{\partial \phi_1} \right)^2 \simeq \frac{h^5 N M_{Pl}^2}{128 \pi^4 m^2} x^2 \left[ (x^2 - 1) \log{ \left( 1 - \frac{1}{x^2} \right)} + (x^2 + 1) \log{ \left( 1+\frac{1}{x^2} \right)} \right]^2,  \\
&\,\\ 
&\eta = M_{Pl}^2 \frac{1}{V} \frac{\partial^2 V}{\partial \phi_1^2} \simeq \frac{h^3 M_{Pl}^2}{8\pi^2 m^2} \left[ (3x^2-1)\log{ \left( 1 - \frac{1}{x^2} \right)} + (3x^2+1)\log{ \left( 1 + \frac{1}{x^2} \right)} \right].
\end{split}
\end{equation}
The slow-roll conditions $\eta \ll 1, \epsilon \ll 1$ are satisfied until $x \sim 1$, provided that $M_{Pl}/m \sim 10^4$ determined to give the correct amplitude of curvature perturbation as we will explicitly show in the next section. Then, the waterfall fields become tachyonic at $x=1$ and settle into the ISS vacuum as we expected. We also have a relation $|\epsilon| \ll |\eta|$ which is used in the next section.

\section{Inflationary predictions}

In this section, we determine the model parameters by comparing our inflationary scenario with the recent cosmological observations. It includes the determination of the mass parameters $m$, $\mu$. In addition, we provide the predictions for the spectral index and the tensor-to-scalar ratio.

First, we consider the initial displacement of the inflaton field $\phi_1^e$. It is obtained by the condition that it can give a sufficient number of e-foldings, $N_e \sim 54 \pm 7$. That is,
\begin{equation}
\begin{split}
N_e &\simeq \frac{1}{M_{Pl}^2} \int_{\phi_1^c}^{\phi_1^e} \left( \frac{V}{V'} \right) d\phi_1 \\
&\simeq 4\pi^2 \left( \frac{m}{M_{Pl}} \right)^2 \frac{x_e^2}{h^3}.
\end{split}
\end{equation}
Then, we can estimate the initial value of the inflaton field such as
\begin{equation}
\phi_e \simeq \frac{\sqrt{N_e N}}{2\pi} h M_{Pl}.
\end{equation}
From this expression, we find a trans-Planckian initial value can be avoided since $\sqrt{N_e N}h < 2\pi$. We do not need to worry that the inflaton may roll off to the lower energy vacua discussed in the previous section when we start from such a large initial value of the inflaton field, because the pseudomoduli field $\phi_2$ is trapped in its expectation value on the ISS vacuum.

Next, we analyse the amplitude of curvature perturbation, which is expressed in terms of the Hubble parameter $H$ and the $\epsilon$ parameter as
\begin{equation}
P_{\zeta}^{1/2} \simeq \frac{1}{\sqrt{2\epsilon}} \left( \frac{H}{2\pi M_{Pl}} \right).
\end{equation}
Inserting the approximate expressions of these parameters, $H=\sqrt{V/3M_{Pl}^2} \simeq \frac{\sqrt{N}m^2}{\sqrt{3}M_{Pl}} $ and $\epsilon \simeq \frac{h^2}{32\pi^2}\frac{N}{N_e}$, we can obtain
\begin{equation}
P_{\zeta}^{1/2} \simeq \frac{1}{h} \sqrt{\frac{4N_e}{3}} \left(\frac{m}{M_{Pl}} \right)^2.
\end{equation}
The WMAP normalization \cite{Spergel:2006hy}, $P_{\zeta}^{1/2} = 4.86 \times 10^{-5}$, implies
\begin{equation}
\frac{m}{M_{Pl}} \sim 10^{-3} \times h^{1/2},
\end{equation}
and hence we obtain $m \sim 10^{14} $ GeV. From the gaugino and scalar mass formulae \eqref{soft}, $m$ is the messenger scale of gauge mediation. The above result shows that the messenger scale is relatively high in our inflationary scenario. When we require TeV-scale gaugino and scalar masses, the SUSY breaking scale is given by $\mu \sim 10^{9.5}$ GeV.

Finally, we provide the observational predictions from our inflationary scenario. The spectral index is given by
\begin{equation}
n_s \simeq 1 + 2\eta \simeq 1 - \frac{1}{N_e} \sim 0.98,
\end{equation}
where we have used the property $|\epsilon| \ll |\eta|$ denoted in the previous section. This red spectrum is consistent with the current observational bound.
In addition, the tensor-to-scalar ratio $r$ is obtained as
\begin{equation}
r = 16\epsilon \simeq \frac{h^2}{2\pi^2} \frac{N}{N_e} \sim 10^{-6}.
\end{equation}
This small value is also consistent with the current bound.

\section{Conclusion}

In this letter, we have proposed an inflationary scenario motivated by the KOO model of direct mediation of low-scale SUSY breaking. This scenario is attractive in that it reconciles both high-scale inflation and gauge mediation in a single sector. Two mass scales are required from the stability of the metastable ISS vacuum. The higher mass scale in the model is determined so that the scenario gives the correct amplitude of curvature perturbation, while the lower mass scale is the SUSY breaking scale to give the order $1$ TeV MSSM soft masses by direct gauge mediation. The messenger scale of gauge mediation is high enough, and hence the Landau pole problem does not occur. As discussed in \cite{Kobayashi:2009rn}, this property is also favorable in light of the little fine-tuning problem occurring in the MSSM.

It is interesting to analyse the reheating process after inflation in our scenario. The analysis is not a trivial work due to the existence of the hidden sector. The gravitino overproduction problem \cite{Endo:2007sz,Weinberg:1982zq,Kawasaki:2008qe} may exist in the present simplest model. We will further elaborate on various issues associated with the present scenario in the upcoming paper \cite{Kamada}.

\section*{Acknowledgments}

We would like to thank K.-I. Izawa, K. Kamada, Y. Ookouchi, O. Seto, T. Shimomura, R. Takahashi, and M. Taki for discussions. This work was supported by the Grant-in-Aid for the Global COE Program "The Next Generation of Physics, Spun from Universality and Emergence" from the Ministry of Education, Culture, Sports, Science and Technology (MEXT) of Japan.

%
%

\end{document}